%% file: main.tex
\documentclass[a4paper,11pt]{article}

\usepackage{pos}
\usepackage{url}

\title{Searches for IceCube Neutrinos Coincident with Gravitational Wave Events}

\ShortTitle{Searches for IceCube Neutrinos Coincident with Gravitational Wave Events}

\author{The IceCube Collaboration \\{\normalsize \normalfont(a complete list of authors can be found at the end of the proceedings)}\\}

\emailAdd{jessie.thwaites@icecube.wisc.edu}
\emailAdd{aswathi.balagopalv@icecube.wisc.edu}
\emailAdd{sahori@wisc.edu}
\emailAdd{mromfoe@wisc.edu}
\emailAdd{acz2122@columbia.edu}

\abstract{
Searches for neutrinos from gravitational wave events have been performed utilizing the wide energy range of the IceCube Neutrino Observatory. We discuss results from these searches during the third observing run (O3) of the advanced LIGO and Virgo detectors, including a low-latency follow-up of public candidate alert events in O3, an archival search on high-energy track data, and a low-energy search employing IceCube-DeepCore. The dataset of high-energy tracks is mainly sensitive to muon neutrinos, while the low energy dataset is sensitive to neutrinos of all flavors. In all of these searches, we present upper limits on the neutrino flux and isotropic equivalent energy emitted in neutrinos. We also discuss future plans for additional searches, including extending the low-latency follow-up to the next observing run of the LIGO-Virgo-KAGRA detectors (O4) and analysis of gravitational wave (GW) events using a high-energy cascade dataset, which are produced by electron neutrino charged-current interactions and neutral-current interactions from neutrinos of all flavors.

\vspace{4mm}
{\bfseries Corresponding authors:}
Jessie Thwaites$^{1*}$, Aswathi Balagopal V.$^{1}$, Sam Hori$^{1}$, M.J. Romfoe$^{1}$, Albert Zhang$^{2}$\\
{$^{1}$ \itshape Dept. of Physics and Wisconsin IceCube Particle Astrophysics Center, University of \\Wisconsin{\textendash}Madison}\\
{$^{2}$ \itshape Columbia Astrophysics and Nevis Laboratories, Columbia University}\\[4mm]
$^*$ Presenter

\ConferenceLogo{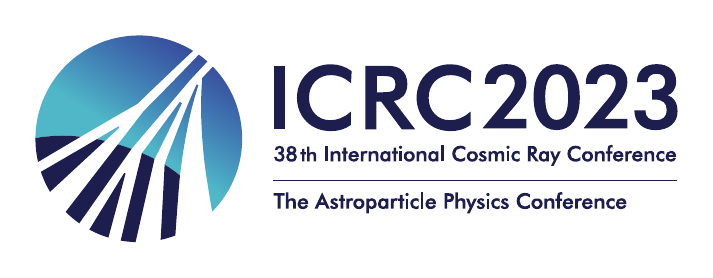}

\FullConference{The 38th International Cosmic Ray Conference (ICRC2023)\\ 26 July -- 3 August, 2023\\ Nagoya, Japan}
}

\begin{document}

\maketitle

\section{Introduction}
Multi-messenger astronomy is a powerful tool that can be used to search for astrophysical sources of high energy neutrinos. By using information from other messengers, we can probe sources of neutrinos that may otherwise be obscured. One such combination that has yet to be observed is a joint source of high energy neutrinos and gravitational waves, despite many previous searches \cite{2020ApJ...898L..10A, 2023ApJ...944...80A, IceCube:2023atb}. 

The third operating run (O3) of the advanced LIGO and Virgo detectors ran from April 1$^\text{st}$, 2019 until March 27$^\text{th}$, 2020. It consisted of a total of 56 detection candidates, which were sent out with low-latency through the Gamma-ray Coordinates Network (now the General Coordinates Network, GCN)\footnote{https://gcn.nasa.gov/}.  After offline analysis, these were released in the second and third gravitational-wave transient catalogs, GWTC-2.1 \cite{2021arXiv210801045T} and GWTC-3 \cite{2021arXiv211103606T}, respectively. The IceCube Collaboration followed up these alerts in real-time, and reported analysis findings. The fourth operating run (O4) of the LIGO, Virgo, and KAGRA (LVK) detectors began on May 24$^\text{th}$, 2023 (although 6 detection candidates were also released during the engineering run, between May 18$^\text{th}$ and May 24$^\text{th}$), and is currently ongoing. The LVK has sent several candidate events over GCN, and IceCube has followed up all significant alerts to date for O4. 

We describe multiple searches for IceCube neutrinos covering a wide range of energies and event selections coincident with gravitational wave candidate sources identified by the LVK. Section \ref{sec:icecube} describes the IceCube Neutrino Observatory, and the three neutrino data selections used in these analyses. Section \ref{sec:realtime} discusses the real-time searches performed in O3 and O4, and section \ref{sec:archival} discusses searches performed using the gravitational wave transient catalogs GWTC-1 \cite{2019PhRvX...9c1040A}, GWTC-2.1 \cite{2021arXiv210801045T}, and GWTC-3 \cite{2021arXiv211103606T}. In section \ref{sec:disc} we provide a discussion of these searches, and a brief outlook.

\section{The IceCube Neutrino Observatory}
\label{sec:icecube}
The IceCube Neutrino Observatory \cite{Aartsen_2017} is a cubic kilometer-scale detector located at the South Pole. It consists of 86 strings with a total of 5160 digital optical modules (DOMs). These DOMs are deployed between 1.45\,km to 2.45\,km below the surface of the ice. For the main array, each string consists of 60\,DOMS, spaced 17\,m apart, with strings placed horizontally 125\,m apart. This spacing in the main array is optimized for high energy (TeV-PeV) neutrinos, with some sensitivity down to energies of roughly~100\,GeV. In the center of the detector is an infill array, called IceCube-DeepCore \cite{IceCube:2011ucd}, which consists of 8 specialized strings, with reduced horizontal spacing of the strings and DOMs placed 7\,m apart on the strings. These strings, as well as the 7~main array strings around them, comprise IceCube-DeepCore, which has sensitivity down to a few GeV. 

\subsection{Event selections}
\label{sec:ev_selections}
\begin{figure}
    \centering
    \includegraphics[width=1.\textwidth]{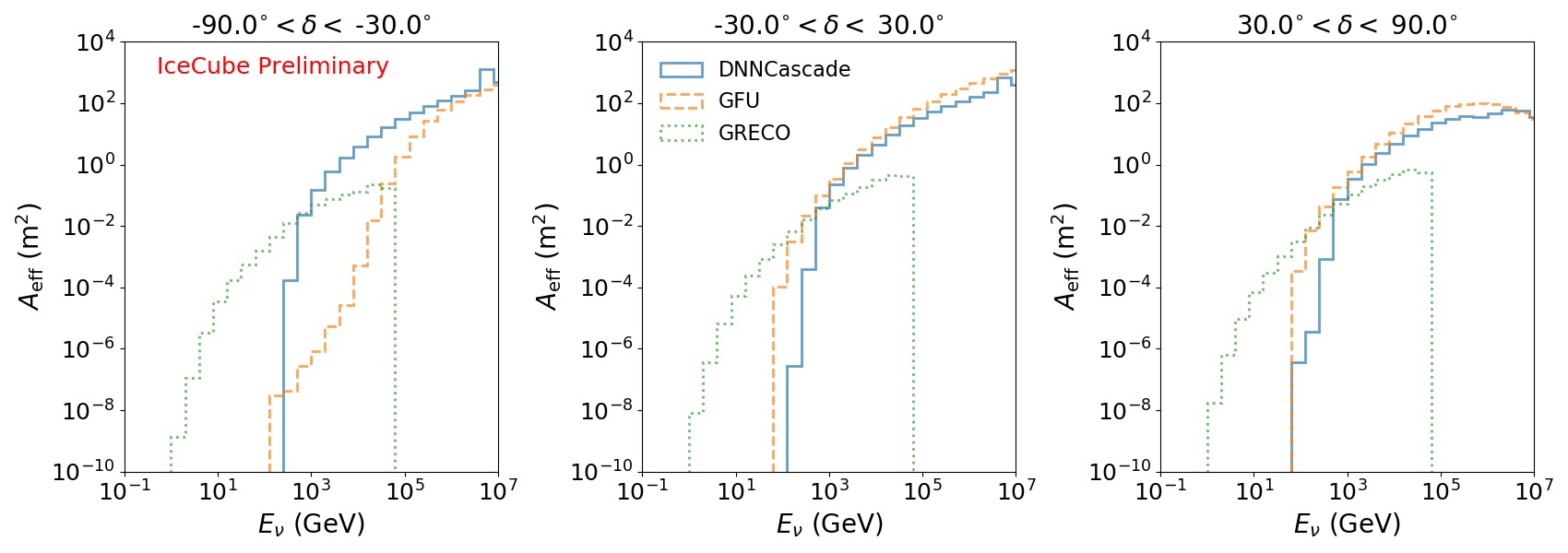}
    \caption{All-flavor effective areas ($\nu+\bar{\nu}$) in different declination bands (Southern sky, Celestial horizon, and Northern sky) for the three data samples discussed here. The GFU sample is only sensitive to $\nu_{\mu}$ CC events while the other datasets are summed across their effective areas for all flavors, including all-flavor NC events as well as $\nu_{e}$ and $\nu_{\tau}$ CC events. These datasets are complimentary to one another; in the Northern sky and at the horizon the GFU dataset provides the best sensitivity, while there is significant improvement with the DNNCascade data sample in the Southern sky. The GRECO dataset is sensitive to a lower energy range, which extends into the low GeV energies.}
    \label{fig:aeff}
\end{figure}

In this work, we describe searches for neutrinos from gravitational wave candidate events using three data samples developed by IceCube. The complimentary effective areas of these selections can be seen in Fig. \ref{fig:aeff}.

At the high energies (above 100 GeV), we use the Gamma-ray Follow-up (GFU) data sample \cite{Aartsen_2016}. This data sample is comprised of track-like events, and is sensitive to muon neutrinos from charged-current interactions interacting in the main array of the IceCube detector. It is available with low latency for use in real-time follow ups of candidate events sent by the LVK. 

At lower energies, we utilize the GeV Reconstructed Events with Containment for Oscillation (GRECO) dataset \cite{IceCube:2022lnv}, which was originally developed for oscillation studies. It uses IceCube-DeepCore, and has sensitivity in the neutrino energy range of a few GeV to some tens of TeV. It is sensitive to neutrinos with all flavors, and both charged- and neutral-current interactions within the DeepCore volume.

We also utilize a new data sample developed by IceCube, which relies on a Deep Neural Network selection of cascade events (DNNCascade sample) \cite{dnncascades}. This data sample shows significant improvement in sensitivity in the Southern sky over the GFU dataset, and provides a complementary search to the GFU search in the high energy regime. It is sensitive to electron neutrino charged-current interactions and neutral-current interactions of neutrinos of all flavors.

\section{Real-time searches using high energy neutrinos in O3 and O4}
\label{sec:realtime}
For the low-latency search for neutrinos from gravitational wave candidate events with IceCube, two pipelines are used. These are an Unbinned Maximum Likelihood (UML) search and the Low-latency Algorithm for Multi-messenger Astrophysics (LLAMA), and both are described in full in \cite{2020ApJ...898L..10A, 2023ApJ...944...80A}. 

The UML search uses the HEALPix pixelization scheme \cite{Górski_2005} to divide the sky into equal area pixels. An all-sky scan is then performed, and a test statistic is calculated for each pixel on the sky. Each pixel is penalized by a spatial weighting term derived from the probability maps published in real-time by the LVK collaboration. The best-fit location is then the pixel on the sky with the maximum TS returned in the scan. The p-value is calculated by comparing this test statistic to a set of background pseudo-experiments. 

The LLAMA search uses a Bayesian odds ratio as a test statistic for the joint gravitational wave and neutrino candidate \cite{Bartos_2017}. The LLAMA search analyzes both low-significance and significant GW candidates from LVK alerts. LLAMA considers information such as the probability of the GW event being astrophysical ($p_{\text{astro}}$) and the luminosity distance of the gravitational wave event in addition to the spatial and temporal overlap of the GW and neutrino candidates. The p-value is obtained by comparing the odds ratio for each event to precomputed background distributions.

\begin{figure}
    \centering
    \includegraphics[width=\textwidth]{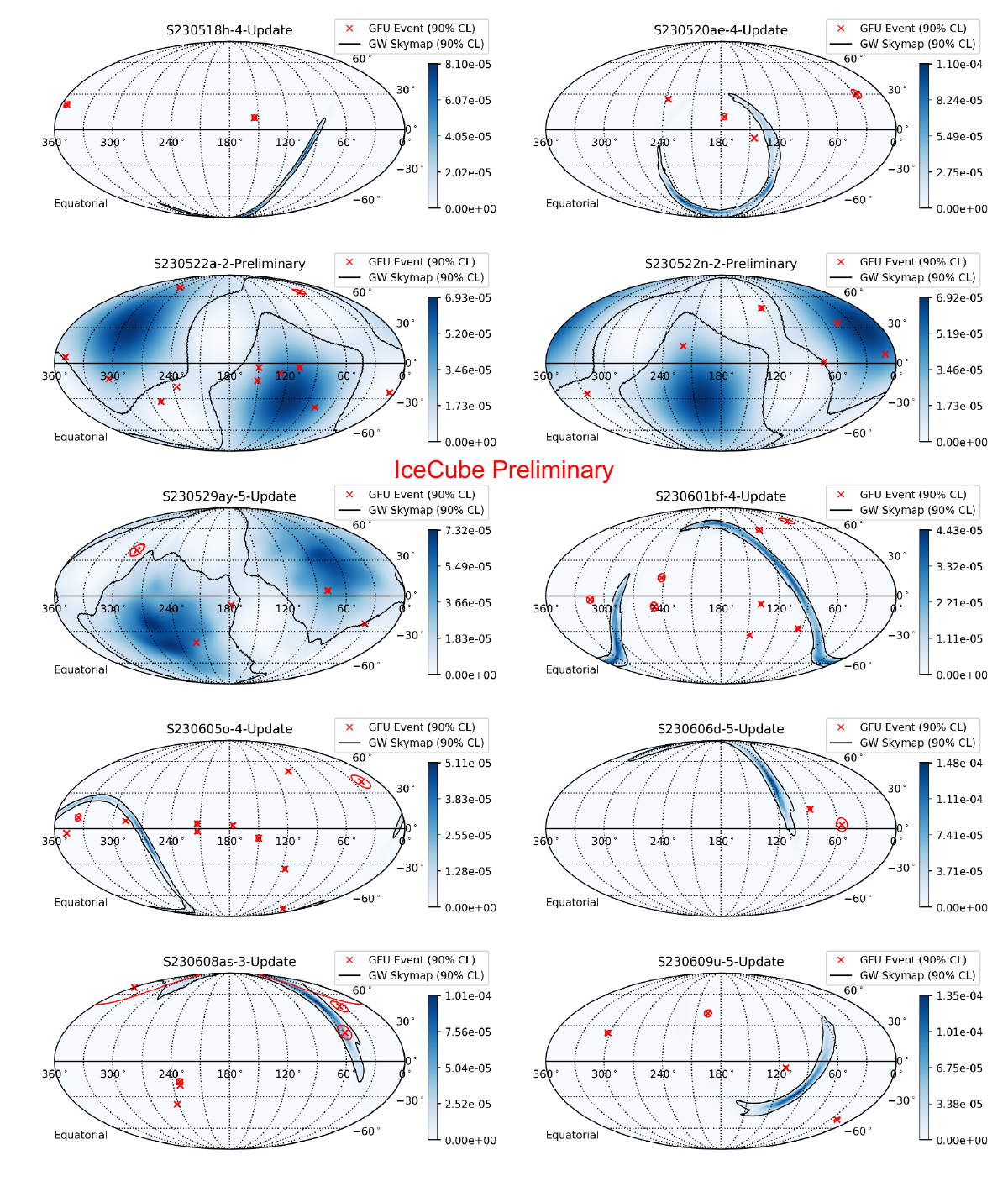}
    \caption{GFU events overlaid on significant gravitational wave probability maps sent by the LVK for O4. Gravitational wave event probabilities are shown in the colorbar, with the black contour showing the 90\% containment of the gravitational wave event map. The event name, sequence number, and type of map is labelled in the title of each panel. Neutrino events are shown in crosses, with the 90\% CL angular uncertainty drawn in circles around each event. Events shown are the first 10 significant event skymaps for O4, including those sent during the engineering run, spanning from May 18th to June 9th, 2023.}
    \label{fig:gallery}
\end{figure}

IceCube has performed follow-ups for all gravitational wave candidate events reported by the LVK. In real-time, multiple skymaps with probabilities are released by the LVK, as more sophisticated processing is applied. These are referred to by their sequence number (an integer which gives the iteration of the GCN that is being followed up) and the map type (Preliminary, Initial, or Update). In O3, results were reported using GCN Circulars with human-in-the-loop vetting \cite{2023ApJ...944...80A}, but because of the increased rate expected during O4, a GCN Notice stream using Kafka has been established\footnote{To subscribe to the alerts, see the documentation at \url{https://gcn.nasa.gov/missions/icecube} and use the topic \emph{gcn.notices.icecube.lvk\_nu\_track\_search}}. Skymaps of the neutrino events observed by IceCube in the on-time window ($\pm$500 seconds around the merger time) for the first alerts with the most recent available skymap for O4 are plotted in Fig. \ref{fig:gallery}. 

\section{Archival searches using the GWTC-1, GWTC-2.1, and GWTC-3 catalogs}

\begin{figure}
    \centering
    \includegraphics[width=0.6\textwidth]{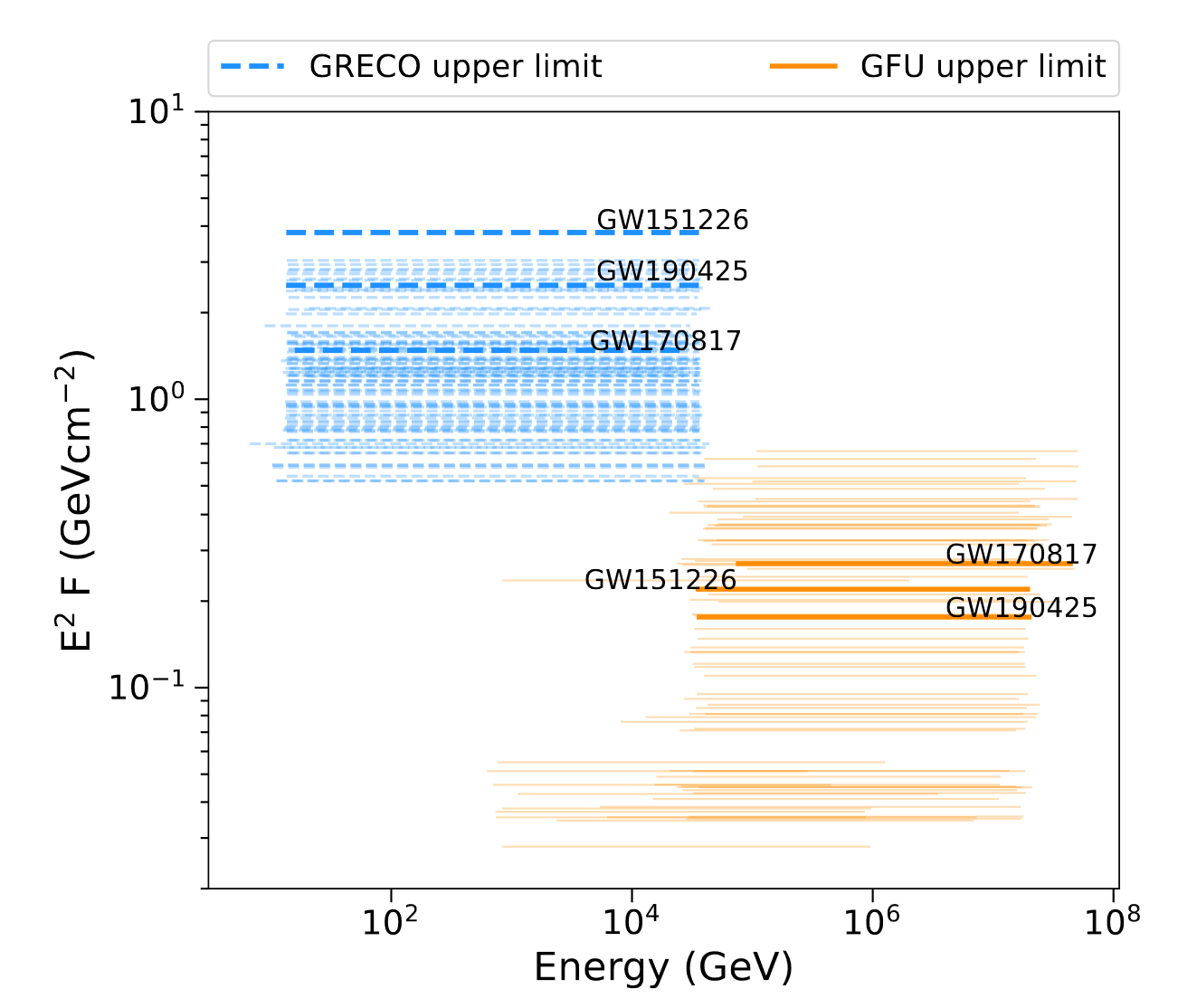}
    \caption{Time-integrated flux upper limits obtained for the 90 GW events obtained in the GRECO (blue dashed) \cite{IceCube:2023atb} and GFU  (orange solid) \cite{2020ApJ...898L..10A, 2023ApJ...944...80A} catalog searches, for a flux with a spectral index of 2.0. The energy ranges shown here are the central 90\% energies contributing to the flux limits at the declinations spanning the 90\% probability regions of the GW skymaps. Three GW events are highlighted here. These are GW151226 (the event with the lowest pre-trial p-value in the GRECO analysis), GW190425 (the only BNS event with a pre-trial p-value < 0.1) and GW170817 (first and only BNS event for which the electromagnetic counterpart has been observed). Figure reproduced from \cite{IceCube:2023atb}.}
    \label{fig:greco-gfu}
\end{figure}

The LIGO-Virgo Collaboration published the confident events observed during the O1 \& O2, O3a and O3b runs in the GWTC-1, GWTC-2.1 and GWTC-3 catalogs respectively (KAGRA also joined for O3b). These catalogs contained both confident as well as marginal GW event classes. Out of these, we used 84 confident BBH, 7 confident NSBH, 1 marginal NSBH and 2 confident BNS events for our analyses.
\label{sec:archival}

\subsection{High energy follow-up: GFU}
Using the GFU data sample, we performed an archival search on confident events published in the GWTC-1, GWTC-2.1 and GWTC-3 catalogs using the same two pipelines (UML and LLAMA) as described in Section \ref{sec:realtime}. The searches were conducted within a 1000~second time window for all GW events. Additionally, we searched for excesses of neutrino emission associated with the neutron-star containing GW events within a 2~week time window using the UML method. We found no significant emission from any of the events analyzed in this study and set 90\% confidence level (CL) upper limits on the neutrino flux for each GW event for all searches. These upper limits for each GW event for the 1000~second time window search are shown in Fig. \ref{fig:greco-gfu}. %We also calculate the upper limits on the isotropic energy emitted in neutrinos ($E_{iso}$), shown in figure \ref{fig:eiso}.

The lowest pre-trial $p$-value observed with both the UML and LLAMA methods was for GW190728\_064510, with values of $4 \times 10^{-2}$ and $8 \times 10^{-3} $ respectively. This event was also one of the candidate-coincident events found in the real-time pipeline, for which the neutrino information was released. We also set upper limits for the isotropic equivalent energy emitted in high-energy neutrinos within the 1000~second time window for each of these merger events. For the 2~week search, we obtained the lowest pre-trial $p$-value of 0.13 for GW200210\_092254. The follow-up with the GFU dataset also included a search for neutrinos coincident with the optical counterpart of GW190521, AGN J124942.3+344929, observed in real-time with the  Zwicky Transient Facility (ZTF) \cite{Graham:2020gwr}. No significant emission was observed with both the UML and LLAMA searches and we report a time-integrated flux-upper limit of  0.081 GeV cm$^{-2}$ and 0.05 GeV cm$^{-2}$ with the UML and LLAMA analyses respectively.

\subsection{Low energy follow-up: GRECO}
The GRECO Astronomy dataset was used to search for low-energy neutrinos associated with binary merger events reported in the catalogs published by LIGO-Virgo. We followed up 83 out of the 84 BBH events, depending on the availability of the GRECO Astronomy dataset. The UML analysis was conducted for the search with a 1000~second time window. We report no significant emission of low-energy neutrinos associated with GW events used for the search. The lowest pre-trial $p$-value of $7.83 \times 10^{-3}$ is reported for the event GW151226 with a time-integrated flux-upper limit of 3.80 GeV cm$^{-2}$. We show the upper limits on the neutrino flux from each event in Fig. \ref{fig:greco-gfu}.

\subsection{Cascade event follow-up: DNNCascades}
\begin{figure}
    \centering
    \includegraphics[width=0.8\textwidth]{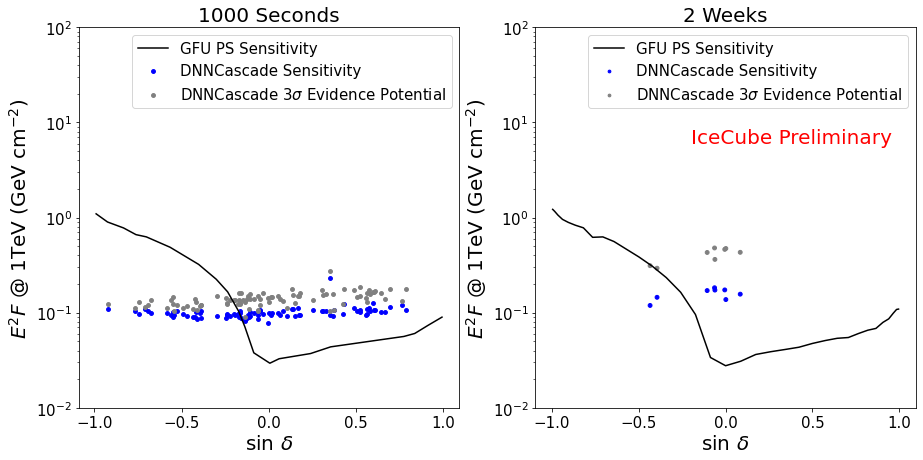}
    \caption{Sensitivity and 3$\sigma$ evidence potential given an injected power law spectrum with a 2.0 index for individual GWTC-2.1 and GWTC-3 gravitational wave candidate events marginalized over each skymap and plotted at the centroid declination compared to the GFU point source sensitivity at different declinations. The 1000~second time window is shown on the left and a 2~week time window on the right. One event in the 1000~second time window has a significantly reduced sensitivity because IceCube livetime only covers about ~50\% of the time window. }
    \label{fig:sens_dnnc}
\end{figure}

We are currently developing an archival analysis of confident events published in GWTC-2.1 and GWTC-3 using an UML methodology with the DNNCascade dataset discussed in section \ref{sec:ev_selections}. The IceCube sensitivities to each gravitational wave event are compared to the GFU track-based searches in Fig. \ref{fig:sens_dnnc}. The smaller background of cascades in the southern hemisphere, and corresponding improved effective area of the dataset shown in Fig. \ref{fig:aeff}, contribute to an order of magnitude improvement in the sensitivity. 

\section{Discussion}
\label{sec:disc}
We present several searches for neutrinos from gravitational wave sources, both in real-time and archival searches using the IceCube Neutrino Observatory. There is also a complementary analysis using extremely low energy events, presented in \cite{Kruiswijk:2023icrc}.

In real-time, the all-sky coverage and small localization of neutrino events relative to gravitational wave candidate event localizations make neutrino follow-ups a powerful tool to identify transients. The increased sensitivity of the LVK detectors and thus increased range provides additional opportunity to search for a joint source of neutrinos and gravitational waves. 

We also present archival searches for neutrinos from GW sources using the catalog of confident gravitational wave events with a time window of 1000~seconds around the merger time \cite{2020ApJ...898L..10A, 2023ApJ...944...80A}. For both of these searches, we do not find evidence of significant emission, and set upper limits on the flux emitted in neutrinos from each GW event. 

We expect near-term improvements to both the neutrino and gravitational wave detectors, which will improve these searches for a joint source. On the gravitational wave detector side, when Virgo is able to rejoin the run we expect to see greatly improved localizations of the events on the sky, which will enhance the capabilities of the searches presented here. The addition of KAGRA will also provide improved triangulation of the events, especially in the next observing run (O5)\footnote{O5 preliminary information, including sensitivites: \url{https://observing.docs.ligo.org/plan/}}. On the neutrino side, the proposed expansions to IceCube in the form of the IceCube Upgrade \cite{Ishihara:2019uL} and IceCube-Gen2 \cite{IceCube-Gen2:2020qha} will improve the sensitivity of these searches. The Upgrade will provide improved energy and direction reconstruction capabilities for neutrinos in the GeV energies and below. For higher energies, the addition of IceCube-Gen2 will increase the instrumented volume of ice, providing an improved sensitivity to TeV-PeV neutrinos. 

In addition, a joint search using both the GRECO and GFU data samples in concert is under development, which synthesizes information from both data samples for use in searching for a joint source of neutrinos and GWs, rather than the separate complementary searches presented here.

% Bibtex references:
\bibliographystyle{ICRC}
\bibliography{references}

\clearpage

\input{authorlist_IceCube.tex}

\end{document}

%% file: authorlist_IceCube.tex
\section*{Full Author List: IceCube Collaboration}

\scriptsize
\noindent
R. Abbasi$^{17}$,
M. Ackermann$^{63}$,
J. Adams$^{18}$,
S. K. Agarwalla$^{40,\: 64}$,
J. A. Aguilar$^{12}$,
M. Ahlers$^{22}$,
J.M. Alameddine$^{23}$,
N. M. Amin$^{44}$,
K. Andeen$^{42}$,
G. Anton$^{26}$
% ...
\\
\\
% ...
$^{12}$ Universit{\'e} Libre de Bruxelles, Science Faculty CP230, B-1050 Brussels, Belgium \\
% ...
$^{17}$ Department of Physics, Loyola University Chicago, Chicago, IL 60660, USA \\
$^{18}$ Dept. of Physics and Astronomy, University of Canterbury, Private Bag 4800, Christchurch, New Zealand \\
% ...
$^{22}$ Niels Bohr Institute, University of Copenhagen, DK-2100 Copenhagen, Denmark \\
$^{23}$ Dept. of Physics, TU Dortmund University, D-44221 Dortmund, Germany \\
% ...
$^{26}$ Erlangen Centre for Astroparticle Physics, Friedrich-Alexander-Universit{\"a}t Erlangen-N{\"u}rnberg, D-91058 Erlangen, Germany \\
% ...
$^{40}$ Dept. of Physics and Wisconsin IceCube Particle Astrophysics Center, University of Wisconsin{\textendash}Madison, Madison, WI 53706, USA \\
% ...
$^{42}$ Department of Physics, Marquette University, Milwaukee, WI, 53201, USA \\
% ...
$^{44}$ Bartol Research Institute and Dept. of Physics and Astronomy, University of Delaware, Newark, DE 19716, USA \\
% ...
$^{63}$ Deutsches Elektronen-Synchrotron DESY, Platanenallee 6, 15738 Zeuthen, Germany  \\
$^{64}$ Institute of Physics, Sachivalaya Marg, Sainik School Post, Bhubaneswar 751005, India \\
% ...

\subsection*{Acknowledgements}

\noindent
The authors gratefully acknowledge the support from the following agencies and institutions:
USA {\textendash} U.S. National Science Foundation-Office of Polar Programs,
U.S. National Science Foundation-Physics Division,
U.S. National Science Foundation-EPSCoR,
Wisconsin Alumni Research Foundation,
Center for High Throughput Computing (CHTC) at the University of Wisconsin{\textendash}Madison,
Open Science Grid (OSG),
Advanced Cyberinfrastructure Coordination Ecosystem: Services {\&} Support (ACCESS),
Frontera computing project at the Texas Advanced Computing Center,
U.S. Department of Energy-National Energy Research Scientific Computing Center,
Particle astrophysics research computing center at the University of Maryland,
Institute for Cyber-Enabled Research at Michigan State University,
and Astroparticle physics computational facility at Marquette University;
Belgium {\textendash} Funds for Scientific Research (FRS-FNRS and FWO),
FWO Odysseus and Big Science programmes,
and Belgian Federal Science Policy Office (Belspo);
Germany {\textendash} Bundesministerium f{\"u}r Bildung und Forschung (BMBF),
Deutsche Forschungsgemeinschaft (DFG),
Helmholtz Alliance for Astroparticle Physics (HAP),
Initiative and Networking Fund of the Helmholtz Association,
Deutsches Elektronen Synchrotron (DESY),
and High Performance Computing cluster of the RWTH Aachen;
Sweden {\textendash} Swedish Research Council,
Swedish Polar Research Secretariat,
Swedish National Infrastructure for Computing (SNIC),
and Knut and Alice Wallenberg Foundation;
European Union {\textendash} EGI Advanced Computing for research;
Australia {\textendash} Australian Research Council;
Canada {\textendash} Natural Sciences and Engineering Research Council of Canada,
Calcul Qu{\'e}bec, Compute Ontario, Canada Foundation for Innovation, WestGrid, and Compute Canada;
Denmark {\textendash} Villum Fonden, Carlsberg Foundation, and European Commission;
New Zealand {\textendash} Marsden Fund;
Japan {\textendash} Japan Society for Promotion of Science (JSPS)
and Institute for Global Prominent Research (IGPR) of Chiba University;
Korea {\textendash} National Research Foundation of Korea (NRF);
Switzerland {\textendash} Swiss National Science Foundation (SNSF);
United Kingdom {\textendash} Department of Physics, University of Oxford.